\documentclass[aps,showpacs,superscriptadress,preprint]{revtex4}
\usepackage{graphicx}

\begin{document}
\title{Improved quark mass density- dependent model with
     quark-sigma meson and quark-omega meson couplings}

\author{Chen Wu$^1$, Wei-Liang Qian$^1$\footnote{wlqian@fudan.edu.cn}
and Ru-Keng Su$^{1,2}$\footnote{rksu@fudan.ac.cn}} \affiliation{
\small 1. Department of Physics, Fudan University,Shanghai 200433, P.R. China\\
\small 2. CCAST(World Laboratory), P.O.Box 8730, Beijing 100080,
P.R. China}

\begin{abstract}
An improved quark mass density- dependent model with the
non-linear scalar sigma field and the $\omega$-meson field is
presented.  We show that the present model can describe saturation
properties, the equation of state, the compressibility and the
effective nuclear mass of nuclear matter under mean field
approximation successfully. The comparison of the present model
and the quark-meson coupling model is addressed.
\end{abstract}

 \maketitle

\section{Introduction}
Owing to the non-perturbative nature of QCD in low energy regions,
phenomenlogical models reflecting the characteristic of the strong
interaction are widely used in the studying of the properties of
hadrons, nuclear matter and quark matters[1-12]. They are based on
different degrees of freedom, for example: nucleons and mesons, or
quarks and gluons, in particular, on hybrid quarks and mesons.
Some of these models  have been proved to be successful. The quark
mass density-dependent model(QMDD)[5] is one of such candidates.

According to the QMDD model, the masses of $u,d$ quarks and
strange quarks (and the corresponding anti-quarks) are given by
\begin{eqnarray}
m_{q} = \frac{B}{3n_{B}}(i = u,d,\bar u,\bar d)   \label{mq1}
\end{eqnarray}
\begin{eqnarray}
m_{s,\bar s } = m_{s0}+\frac{B}{3n_{B}}     \label{mq2}
\end{eqnarray}
where $n_{B}$ is the baryon number density, $m_{s0}$ is the
current mass of the strange quark, and  $B$ is the bag constant.
At zero temperature
\begin{eqnarray}
n_B=\frac{1}{3}(n_u+n_d+n_s),
\end{eqnarray}
where $n_u,n_d,n_s$ represent the density of u quark, d quark, and
s quark, respectively. The basic hypothesis Eqs.(1) and (2)
corresponds to a quark confinement mechanism because if quark goes
to infinite space, the volume of the system tends to infinite,
$n_B$ approachs  to zero and the $m_q$ goes to infinite, and the
infinite quark mass prevents the quark from going to infinite. The
confinement mechanism is similar to that of the MIT bag model.

Although the QMDD model can provide a description of confinement
and explain  many  dynamical properties of strange quark matter,
but it is still an ideal quark gas model and cannot explain the
temperature T vs. density $\rho$ deconfinement phase diagram of
QCD and the properties of nuclear matter[13-14]. To overcome this
difficulty, we have introduced a coupling between quark and
nonlinear scalar field to improve the QMDD model in a previous
paper[15]. We have found  the wave functions of the ground state
and the lowest one-particle excited states. By using these wave
functions, we calculated many physical quantities such as
root-mean-squared radius, the magnetic moment
 of nucleon to compare with experiments  and come to a
conclusion that this improved QMDD model is successful to explain
the properties of nucleon. In ref[16], we extended this model to
finite temperature and studied its soliton solution by means of
the finite temperature quantum field theory. The critical
temperature of quark deconfinement $T_C$ and the
temperature-dependent bag constant $B(T)$ are found. The results
of improved QMDD(IQMDD) model are qualitatively similar to that
obtained from Freidberg-Lee soliton bag model[17, 18].

Instead of studying the nucleon properties, we hope to employ the
IQMDD to investigate the physical properties of nuclear matter in
this paper. As was shown by the Walecka model[1] and the QMC
model[7-10] early, a neutral vector field coupled to the conserved
baryon current is very important for describing bulk properties of
nuclear matter. The large neutral scalar and vector contributions
have been observed empirically from NN scattering amplitude. The
main qualitative features of the nucleon-nucleon interaction: a
short range repulsion between baryons coming from $\omega $-meson
exchange, and a long-range attraction between baryons coming from
$\sigma$-meson exchange must be included in a successful model.
Obviously, if we hope to employ the IQMDD model to mimick this
repulsive and attractive interactions, except the quark and
$\sigma$-meson interaction, the $\omega$ meson and the $qq\omega$
coupling must be added. This motivate us to introduce $\omega$
mesons and the $qq\omega$ coupling in the IQMDD model in this
paper. In this new IQMDD model, the nonlinear scalar field
coupling with quarks  forms a soliton bag, and the $qq\omega$
vector coupling gives the repulsion between quarks. We will prove
that this model can give us a successful description of nuclear
matter.

The organization of this paper is as follows. In the next section,
we give the main formulae of the IQMDD model under the mean field
approximation at zero temperature. In the third section, some
numerical results are contained. The last section contains a
summary and discussions.

\section{Formulae of the improved QMDD model}
 The Lagrangian density of the IQMDD model is :
 \begin{eqnarray}
L =
\bar{\psi}[i\gamma^{\mu}\partial_\mu-m_{q}+g^q_\sigma\sigma-g^q_\omega\gamma^\mu\omega_\mu
]\psi+\frac{1}{2}\partial_{\mu}\sigma\partial^{\mu}\sigma-U(\sigma)
-\frac{1}{4}F_{\mu\nu}F^{\mu\nu}+\frac{1}{2}m_{\omega}^2\omega_\mu\omega^\mu
\end{eqnarray}
where
\begin{eqnarray}
F_{\mu\nu}=\partial_\mu \omega_\nu-
\partial_\nu \omega_\mu
\end{eqnarray}
and the quark mass $m_q$ is given by Eqs.(1) and (2),
 $m_\sigma$
and $m_\omega$ are the masses of $\sigma$ and $\omega$ mesons,
$g_\sigma^q$ and $g_\omega^q$ are the couplings constant between
quark-$\sigma$ meson and quark-$\omega$ meson respectively. And
\begin{eqnarray}
 U(\sigma) = \frac{1}{2}m_\sigma^2\sigma^2 +
\frac{1}{3}b\sigma^3+ \frac{1}{4}c\sigma^4+B
\end{eqnarray}
\begin{eqnarray}
-B=\frac{m_\sigma^2}{2}
\sigma_v^2+\frac{b}{3}\sigma_v^3+\frac{c}{4}\sigma_v^4.
\end{eqnarray}
where $\sigma_v$ is the absolute minimum of $U(\sigma),
U(\sigma_v)=0$ and $U(0)=B$.

 It can easily show the equation of motion for quark field in
the whole space is
\begin{eqnarray}
[\gamma^\mu(i\partial_\mu+g_\omega^q \omega_\mu)-(m_q-g_\sigma^q
\sigma)]\psi=0
\end{eqnarray}
Under mean field approximation, the effective quark mass $m_q^{*}$
is given by:
\begin{eqnarray}
 m_q^{*}=m_q-g_\sigma^q\bar{\sigma}
\end{eqnarray}
In nuclear matter, three quarks constitute a bag, and  the
effective nucleon mass is obtained from the bag energy and reads:
\begin{eqnarray}
 M_N^* = \Sigma_q E_q =\Sigma_q \frac{4}{3}\pi R^3
\frac{\gamma_q}{(2\pi)^3}\int_0^{K_F^q}\sqrt{{m^*_q}^2+k^2}
(\frac{dN_q}{dk})dk
\end{eqnarray}
where quark degeneracy $\gamma_q$=6, $K_F^q$ is the Fermi energy
of quarks. $dN_q/dk$ is the density of states for various quarks
in a spherical cavity. It is given by[19]:
\begin{eqnarray}
 N(k) = A(kR)^3+B(KR)^2+C(KR)
\end{eqnarray}
where
\begin{eqnarray}
   A=\frac{2\gamma_q}{9\pi}.
\end{eqnarray}
\begin{eqnarray}
   B(\frac{m_q}{k})=\frac{\gamma_q}{2\pi}\{[1+(\frac{m_q}{k})^2]arctan(\frac{k}{m_q})-\frac{m_q}{k}-\frac{\pi}{2}
   \}.
\end{eqnarray}
\begin{eqnarray}
   C(\frac{m_q}{k})=\tilde{C}(\frac{m_q}{k})+
   (\frac{m_q}{k})^{1.45}\frac{\gamma_q}{3.42(\frac{m_q}{k}-6.5)^2+100}
   .
\end{eqnarray}
\begin{eqnarray}
   \tilde{C}(\frac{m_q}{k})=\frac{\gamma_q}{2\pi}\{\frac{1}{3}+
   (\frac{m_q}{k}+\frac{k}{m_q})arctan(\frac{k}{m_q})-\frac{\pi
   k}{2m_q}\}.
\end{eqnarray}
Eqs. (12) and (13) are in good agreement with those given by
multireflection theory[21, 22] and the Eqs. (14) and (15) are
given by a best fit of numerical calculation for the MIT bag
model. The curvature term $\tilde{C}$ cannot be evaluated by this
theory except for two limiting cases $m_q\rightarrow 0$ and
$m_q\rightarrow \infty$. Madsen[20] proposed the Eq. (15), but as
was pointed out by Ref. [19], the best fit of numerical data is
given by Eq. (14). This density of state has been employed by
Refs. [13,14] to study the strangelets.

The Fermi energy $K_F^q$  of quarks is given by
\begin{eqnarray} 3= \frac{4}{3}\pi R^3 n_B
\end{eqnarray}
where $n_B$ satisfies
\begin{eqnarray} n_B =\Sigma_q \frac{ \gamma_q}{(2\pi)^3} \int_0^{K_F^q}(\frac{dN_q}{dk}) dk
\end{eqnarray}
The bag radius $R$ is determined by the equilibrium condition for
the nucleon bag:
\begin{eqnarray} \frac{\delta M^*_N}{\delta R}=0
\end{eqnarray}

In nuclear matter, the total energy density is given by
\begin{eqnarray}
\varepsilon_{matter}= \frac{\gamma_N}{(2\pi)^3} \int_0^{K_F^N}
\sqrt{{M_N^*}^2+p^2} dp^3 +
\frac{g_\omega^2}{2m_\omega^2}\rho_B^2+\frac{1}{2}m_\sigma^2
\bar\sigma^2+\frac{1}{3}b\bar\sigma^3+\frac{1}{4}c{\bar\sigma}^4
\label{e}
\end{eqnarray}
where $\gamma_N=4$ is degeneracy of nucleon, $K_F^N$ is fermi
energy of nucleon and $\rho_B$ is the density of nuclear matter
\begin{eqnarray}
 \rho_B =\frac{ \gamma_N}{(2\pi)^3} \int_0^{K_F^N} d^3 k
\end{eqnarray}
In Eqs. (19), $g_\omega$ is the coupling constant between the
nucleon and the $\omega$ meson and it satisfies
$g_\omega=3g_\omega^q$. As that of the QMC model[7], the
$\bar\sigma$ is yielded by the equation:
\begin{eqnarray}
m_\sigma^2 \bar\sigma+b\bar\sigma^2+c\bar\sigma^3 = -
\frac{\gamma_N}{(2\pi)^3} \int_0^{K_F^N}
\frac{M_N^*}{\sqrt{{M_N^*}^2+p^2}} d^3 p (\frac{\partial
M_N^*}{\partial \bar\sigma})_R
\end{eqnarray}

Eqs. (9)-(21) form a complete set of equations and we can solve
them numerically. Our numerical results will be shown in the next
section.

\section{numerical result}
Before numerical calculation, let us consider the parameters in
IQMDD model. As that of Ref.[1, 23], the masses of $\omega$-meson
and $\sigma$-meson are fixed as $m_\omega=783$ MeV, $m_\omega=509$
MeV respectively. We choose the bag constant $B=174$ MeV fm$^{-3}$
to fit the  mass of nucleon  $M_N=939$ MeV. When $B$ is
determined, the parameters $b$ and $c$ are not independent because
of Eq. (7). we choose the b is free parameter. There are still
three parameters, namely, $ g_\omega^q , g^q_\sigma, b$ are needed
to be fixed in IQMDD model.

To study the physical properties of nuclear matter, we investigate
 the nuclear saturation, the equation of state and the
 compressibility. The pressure of nuclear matter P is given by
\begin{eqnarray}
 P =\rho_B^2 \frac{\partial}{\partial \rho_B} \frac{\varepsilon_{matter}}{\rho_B}
\end{eqnarray}
where $\rho_B$ is the baryon density. The compressibility for
nuclear matter reads:
\begin{eqnarray}
 K = 9\frac{\partial}{\partial \rho_B}P
\end{eqnarray}
at saturation point, the binding energy per particle $E/A= -15$
MeV, and the saturation density $\rho_0=0.15$ fm$^{-3}$.

Our numerical results are shown in Fig. 1-Fig. 4. In Fig. 1, we
choose  $\omega$-meson and $\sigma$-meson satisfy $\bar\sigma=0,
\bar\omega=0$ and depict the bag energy as a function of bag
radius at zero temperature. We find the stable radius of a "free"
nucleon $R=0.85$ fm.

In Figs. 2-4 we show the effective mass $M^\star$ of nucleon, the
saturation curve and the equations of state of nuclear matter at
zero temperature for IQMDD model respectively, where we fix the
parameter b=-1460 (MeV), $g_\sigma=4.67$ and $g_\omega=2.44$
respectively, and find $E/A=-15$ MeV and $\rho_0=0.15$ fm$^{-3}$
and $K(\rho_0)=210$ MeV. We find our model can explain the
properties of nuclear matter successfully.

To illustrate our results more transparently, we show the
dependence of the properties of nuclear matter on the parameters
$b, g_\sigma^q, g_\omega^q$ in Table.1 for fixing binding energy
$E/A=-15$ MeV and $\rho_0=0.15$ fm$^{-3}$. We find that the
compressibility $K(\rho_0)$ and effective nucleon mass
$M^*_N(\rho_0)$ at saturation point all decrease when $
g_\sigma^q, g_\omega^q$ increase and b decreases. At was shown in
Table.1, the variational regions for  $K(\rho_0)$ and
$M^*_N(\rho_0)$ are small, and the decreasements of $K(\rho_0)$
and $M^*_N(\rho_0)$ are slowly.
\\
\begin{tabular}
{p{1.7cm}p{2.0cm}p{2.0cm}p{2.5cm}p{2.5cm}}
\multicolumn{5}{c}{TABLE 1.  Variation of the nuclear matter
properties   to b.   } \\
\hline\hline  b(MeV)   &$g_\sigma^q$ &$g_\omega^q$
 &K($\rho_0$)(MeV)   &$M^*_N(\rho_0)$(MeV)
\\\hline
-800 &4.59  &2.35     &218.8 &782.9  \\
-1000 &4.61  &2.38     &215.5 &781.2  \\
-1200 &4.64  &2.40     &213.6 &778.5  \\
-1400 &4.66  &2.43     &211.2 &776.8  \\
-1600 &4.69  &2.46     &208.1 &774.3  \\
-1800 &4.71  &2.48     &205.7 &772.6  \\
\hline\hline
\end{tabular}

It was pointed in  the Refs. [24] early, adding  a nonlinear
scalar field in the model will cause unphysical behavior under
mean field approximation in nuclear matter. This can  easily be
seen from Eq.(21) because the left hand side of Eq. (21) is a
cubic order function of $\bar\sigma$, and $\bar\sigma=0$ is one of
its solutions. There are two solutions in low-density regions. In
Fig. 5 and Fig. 6, these two solutions are shown explicitly for
$\bar\sigma$ vs. $\rho_B$ curve and for $M_N^\star$ vs. $\rho_B$
curve respectively where the parameters are fixing as $b=-3655$
(MeV). Noting that the term of non-linear scalar field is
essential to form a soliton bag, we conclude that the unphysical
branch cannot be avoided  for the soliton solution under mean
field approximation. Fortunately, the lower branch cannot be ended
at the point($M_N=939$ MeV, $\rho_B=0$), and give us a
experimental value of nucleon mass, we will give up this
unphysical branch in our calculation.

 Finally, It is of interest to compare the properties of nuclear
 matter for  IQMDD model and for the QMC model. Our results are
 shown in Table. 2. we find their results are very similar. But as
 was pointed in our previous paper[15], the first advantage of the
 IQMDD model is that the MIT bag boundary constraint has been
 given up because it mimicks to a Friedberg-lee soliton bag
 model[13-16]. The second advantage of the IQMDD model is that the
 interactions between $qq\sigma$ and $qq\omega$ are extended to the
 whole free space. We can easily write  down the propagators of
 quarks, $\sigma$-meson and $\omega$-meson respectively and do the
 many-body calculations beyond mean field approximation. But for
 QMC model, the propagators of quarks, $\sigma$-meson and
 $\omega$-meson cannot be written down easily because one must
 consider the multireflection by the MIT bag boundary as well as
 the effect of the interactions limited into a nucleon space
 only. The IQMDD model provides a good substitute of the QMC model
 which is more suitable for the study of nuclear matter beyond
 mean field.
\\
\begin{tabular}
{p{3.7cm}p{2.0cm}p{2.0cm}p{2.0cm}cc}
\multicolumn{6}{c}{TABLE 2.  Comparison of properties for IQMDD and QMC model.  } \\
\hline\hline  &R(fm) & ${g_\sigma^q}$ &${g_\omega^q}$
 &K($\rho_0$)(MeV)   &$M^*_N(\rho_0)$(MeV)
\\\hline
QMC &0.80  &5.53  &1.26  &200 &851   \\
IQMDD(b=0) &0.85  &4.54  &2.21  &227 &798 \\
IQMDD(b=-1460) &0.85  &4.67  &2.44  &210 &775\\
QHD-1 &  &$g_\sigma=9.57$  &$g_\omega=11.6$  &540 &522\\
 \hline\hline
\end{tabular}
\\

\section{Summary and discussion}
In summary, we present an Improved quark mass density dependent
model which has the non-linear $\sigma$ meson field, and the
$\omega$ meson field. The $qq\sigma$ coupling and the $qq\omega$
coupling are introduced to mimick the repulsive and the attractive
interactions between quarks in this model. It is shown that the
present model is successful for describing the saturation
properties, the equation of state and compressibility of nuclear
matter. The effective  nucleon mass decreases with baryon density
in this model more rapidly than that of QMC model. After comparing
the IQMDD model and the QMC model, we come to a conclusion that
the IQMDD model is a good substitute for QMC model.

\begin{figure}[tbp]
\includegraphics[width=14cm,height=20cm]{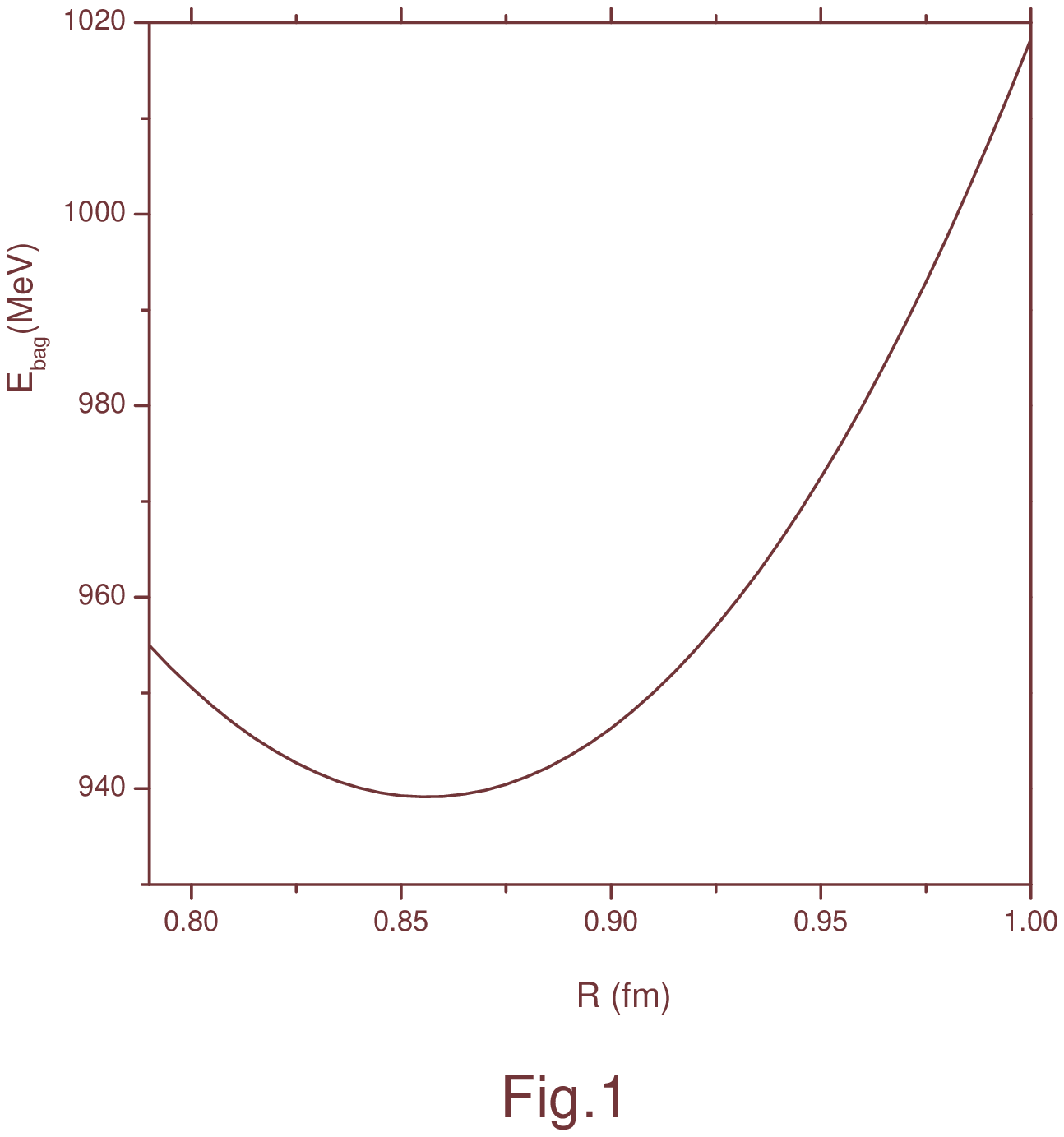}
\caption{The bag energy as a function of bag radius at zero
temperature for $\bar\omega=0, \bar\omega=0$. }

\end{figure}

\begin{figure}[tbp]
\includegraphics[width=14cm,height=20cm]{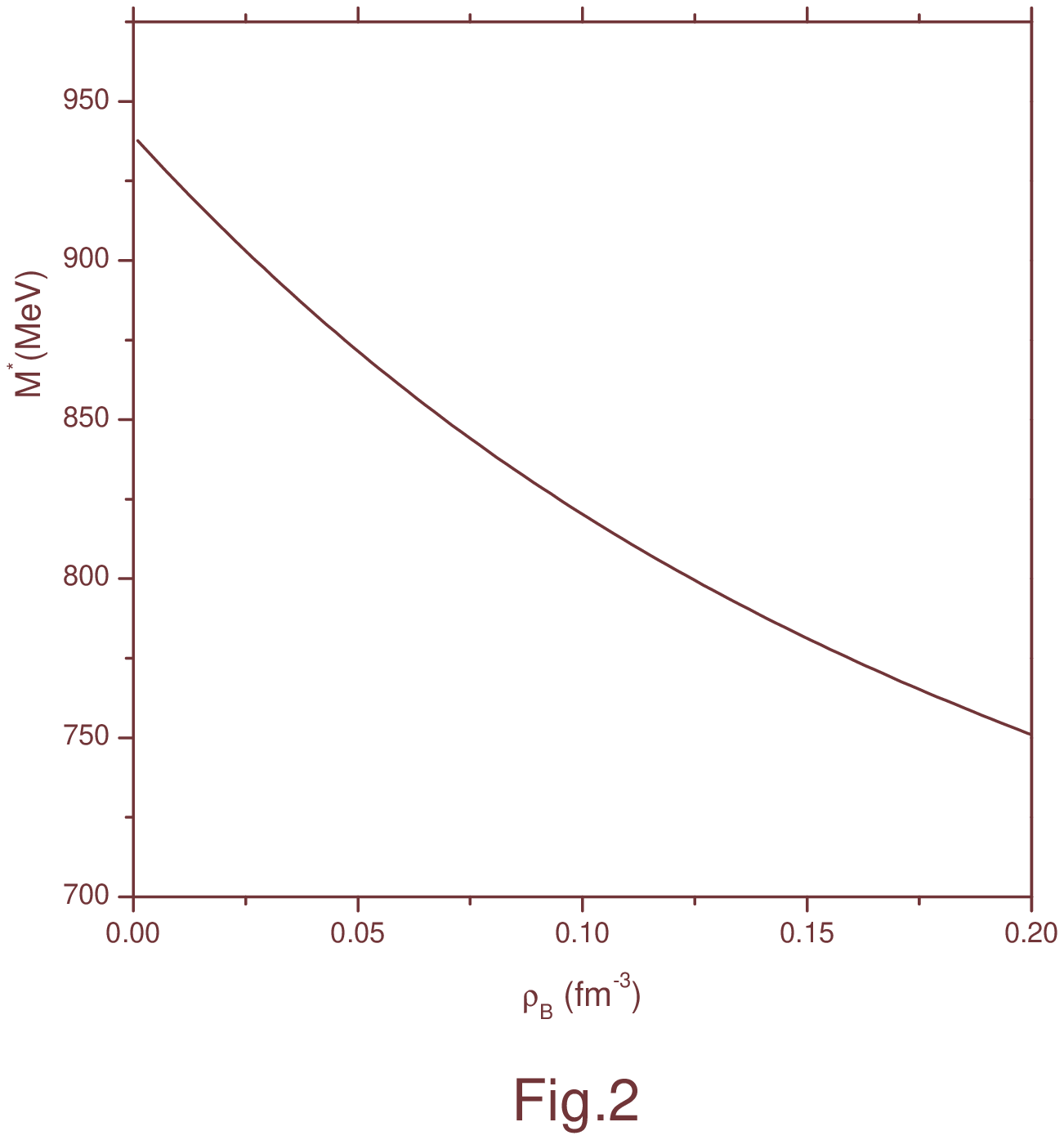}
\caption{Effective nucleon mass vs. baryon density at zero
temperature where the parameters $g_\sigma^q=4.67,
g_\omega^q=2.44, b=-1460$ (MeV).}
\end{figure}

\begin{figure}[tbp]
\includegraphics[width=14cm,height=20cm]{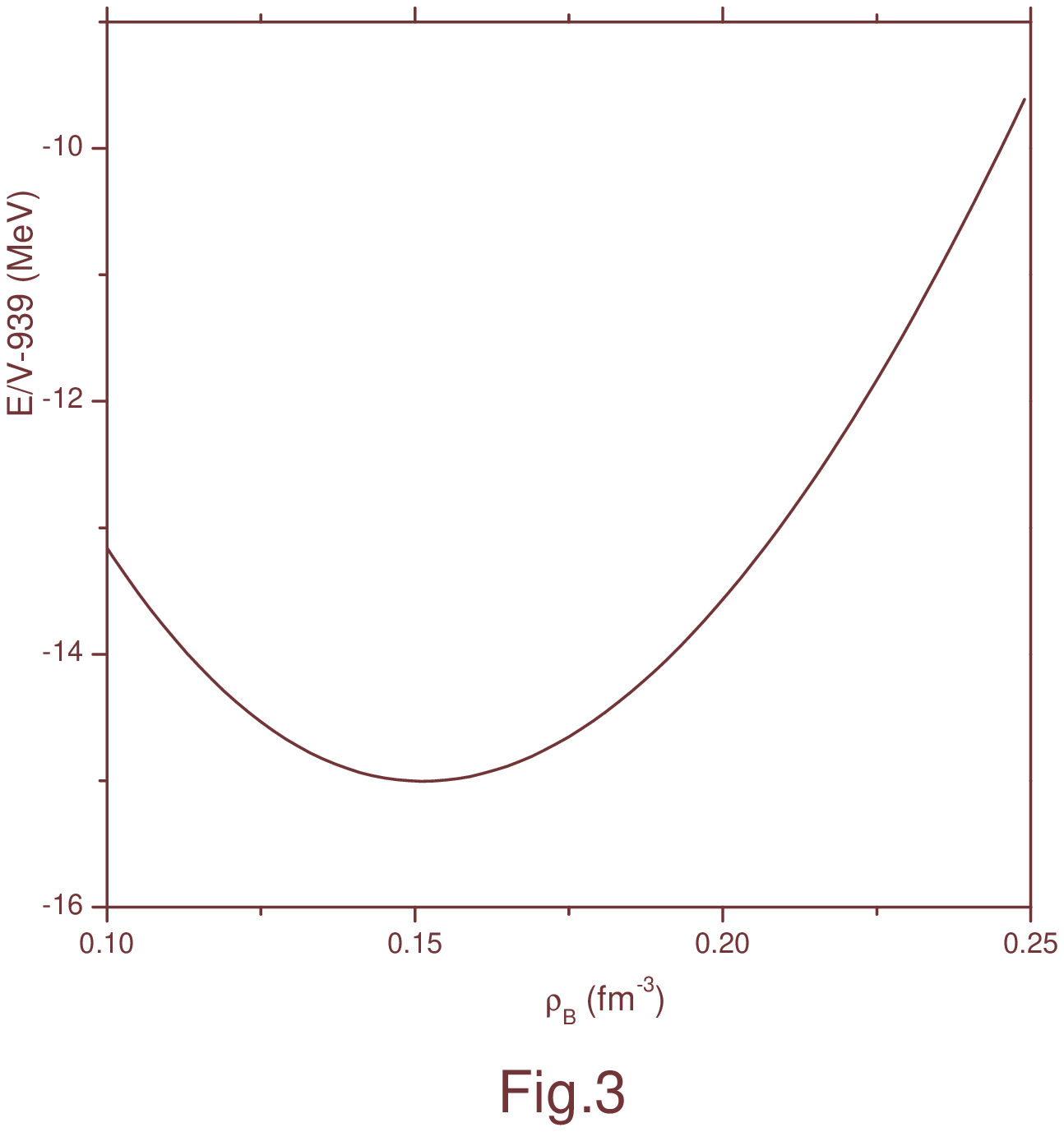}
\caption{Saturation curve of nuclear matter at zero temperature.
the parameters is same as that of Fig. 2.}
\end{figure}

\begin{figure}[tbp]
\includegraphics[width=14cm,height=20cm]{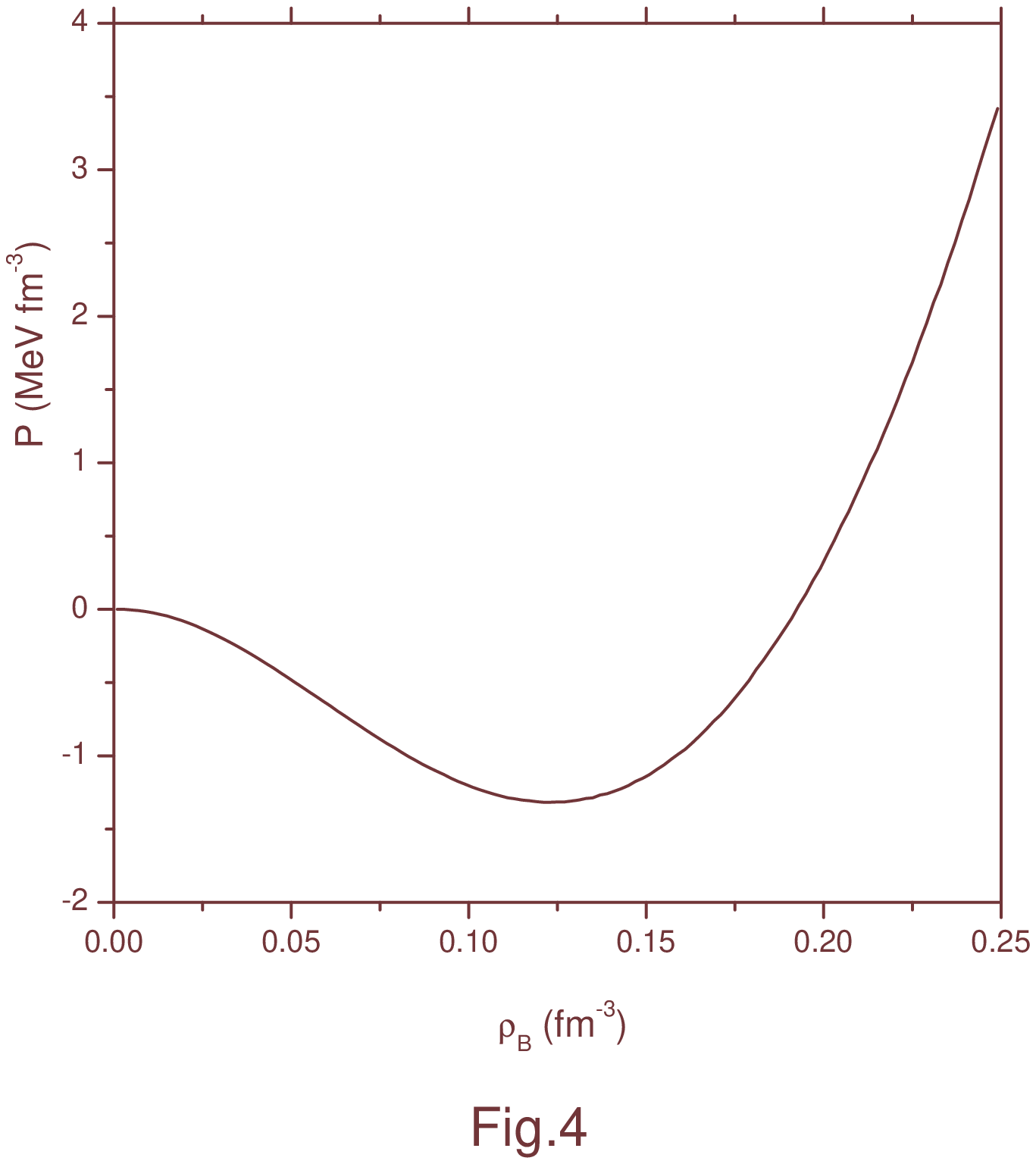}
\caption{Pressure of nuclear matter as a function of $\rho_B$. the
parameters is same as that of Fig. 2.}
\end{figure}

\begin{figure}[tbp]
\includegraphics[width=14cm,height=20cm]{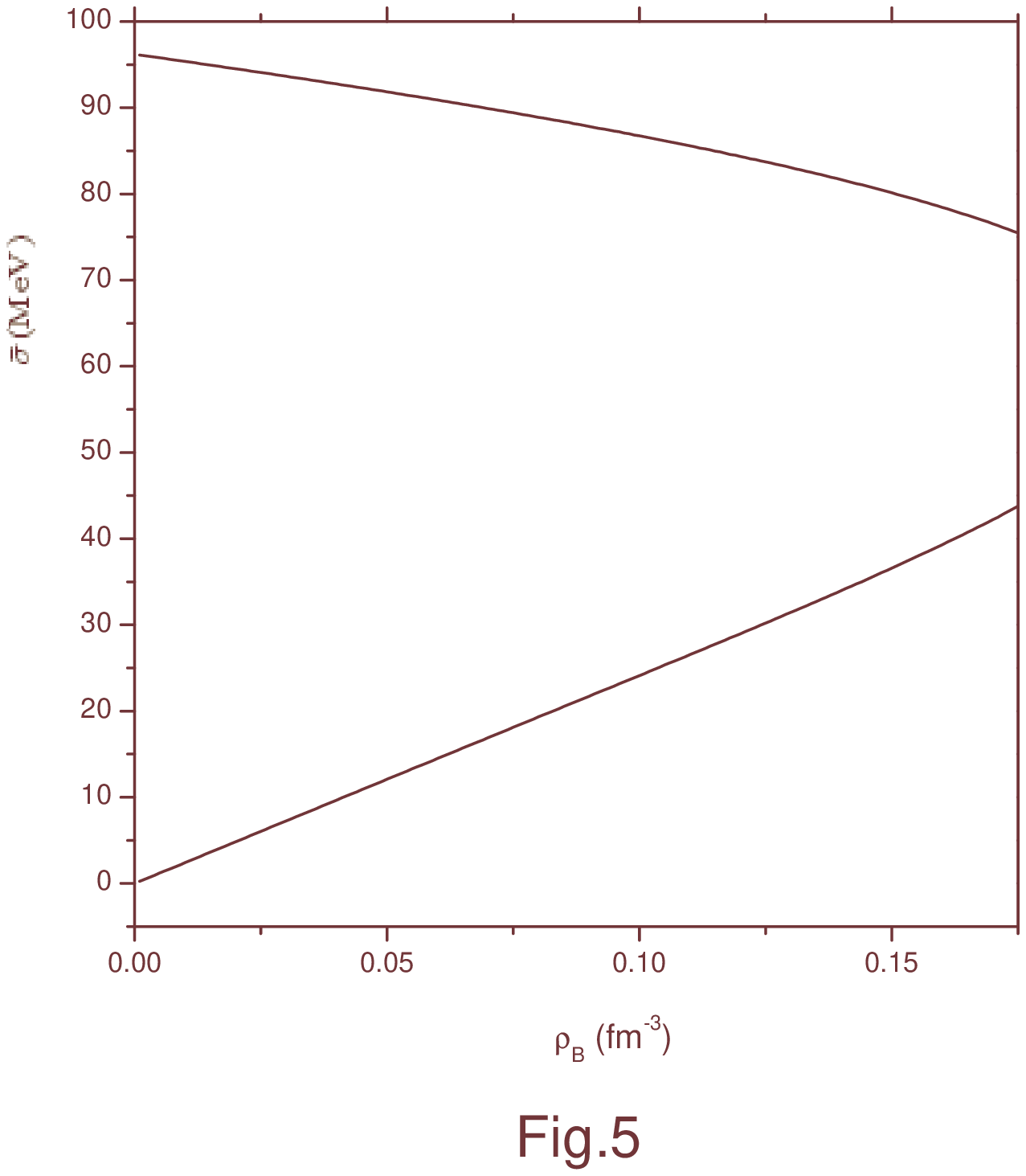}
\caption{the $\bar\sigma$ field  vs. baryon density  for b=-3655
(MeV), $g_\sigma^q=5.23, g_\omega^q=3.12$.}
\end{figure}

\begin{figure}[tbp]
\includegraphics[width=14cm,height=20cm]{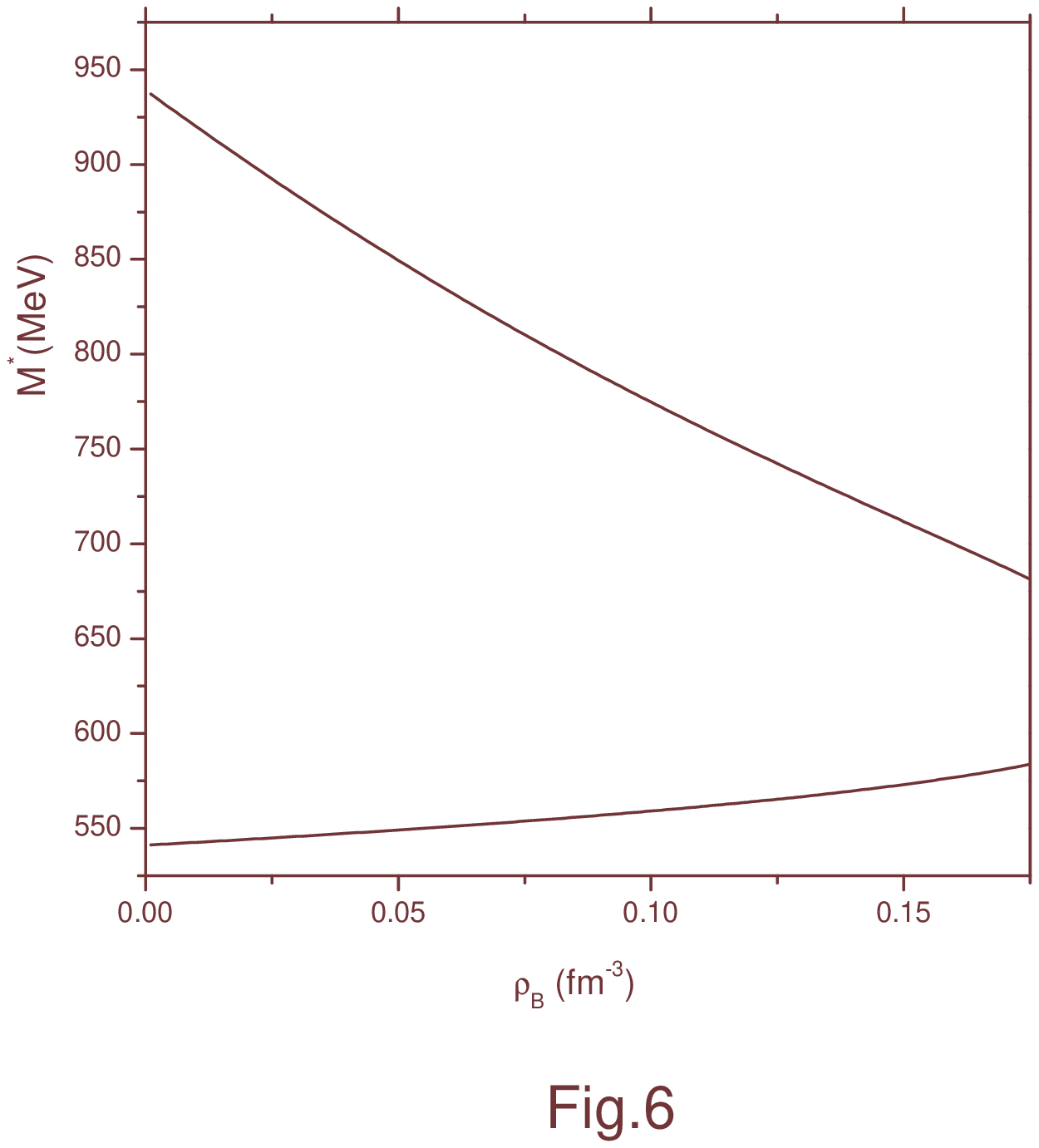}
\caption{Effective nucleon mass $M^\star$ vs. baryon density. the
parameters is same as that of Fig. 5. }
\end{figure}
\end{document}